\documentstyle[12pt,epsfig]{article}
                                                                                
\renewcommand{\thepage}{}

\begin{document}                                                                
\date{}
                                                                                
\title{                                                                         
{\vspace{-3cm} \normalsize                                                      
\hfill \parbox{50mm}{DESY 96-047}}\\[25mm]
Violations of universality                                   \\
in a vectorlike extension of the standard model              \\[12mm]}
\author{ I. Montvay                                          \\
Deutsches Elektronen-Synchrotron DESY,                       \\
Notkestr.\,85, D-22603 Hamburg, Germany}                                       
                                                                                
\newcommand{\be}{\begin{equation}}                                              
\newcommand{\ee}{\end{equation}}                                                
\newcommand{\half}{\frac{1}{2}}                                                 
\newcommand{\rar}{\rightarrow}                                                  
\newcommand{\lar}{\leftarrow}                                                   
                                                                                
\maketitle                                                                      
                                                                                
\begin{abstract} \normalsize
 Violations of universality of couplings in a vectorlike extension of
 the standard model with three heavy mirror fermion families are
 considered.
 The recently observed discrepancies between experiments and the
 standard model in the hadronic branching fractions $R_b$ and $R_c$
 of the Z-boson are explained by the mixing of fermions with their
 mirror fermion partners.
\end{abstract}       

\newpage
\renewcommand{\thepage}{\arabic{page}}
\section{Introduction}\label{sec1}
 The non-perturbative definition of chiral gauge theories in lattice
 regularization encounters great difficulties \cite{LATTICE,ROMA}.
 In fact, up to now no acceptable path-integral formulation with
 exact gauge invariance at finite cut-off is known.
 The basic obstacle is ``fermion doubling'', which occurs under very
 general circumstances \cite{NINI}.
 Therefore, it appears natural to look for a formulation based on
 exact vectorlike gauge invariance \cite{GAUINV}, which requires the
 doubling of the light fermion spectrum \cite{FAMILIES} by mirror
 fermions \cite{MIRROR}.
 In this vectorlike extension of the standard model the chiral
 asymmetry of the light fermion spectrum is a low energy phenomenon.
 At high energies, above the scale of the vacuum expectation value of
 the Higgs scalar field, the theory becomes symmetric.

 A basic feature of the vectorlike extension of the standard model
 is the mixing of the low-lying fermion states with their heavy mirror
 fermion partners.
 The non-zero mixing angles appear in the couplings to the gauge vector
 bosons and result, among other things, in some small breaking of the
 universality.
 In fact, some universality relations are known to be fulfilled to a
 good accuracy.
 In order to reproduce them, together with some other basic facts of
 electroweak phenomenology, one has to choose a particular mixing
 scheme \cite{FAMILIES,CSIFOD}.
 This mixing scheme still leaves some room for the breaking of
 universality at some other places.
 In particular, as it will be shown in this paper, the discrepancies
 between experiments and the standard model in the hadronic branching
 fractions $R_b$ and $R_c$ of the Z-boson can be explained.

 Before discussing a simple viable choice of mixings in section
 \ref{sec3}, first, in the next section, the general properties of
 fermion mirror fermion mixing will be shortly summarized.

\section{The fermion mirror fermion mixing}\label{sec2}
 The mirror partners of light fermions have the same
 $\rm SU(3) \otimes SU(2)_L \otimes U(1)_Y$ quantum numbers but
 opposite chiralities.
 For instance, the right-handed chiral components of mirror leptons             
 form a doublet with $Y=-1$ with respect to
 $\rm SU(2)_L \otimes U(1)_Y$.
 The mirror partners must be heavy enough in order to forbid their
 unobserved pair production at present accelerators.
 Their first observable consequences follow from the mixing with light
 fermions.
 The phenomenologically acceptable mixing schemes were discussed
 in ref.~\cite{FAMILIES} and will be shortly repeated here for the
 reader's convenience.

 In order to discuss the mixing schemes, let us first consider the              
 simplest case of a single fermion ($\psi$) mirror fermion ($\chi$)             
 pair.                                                                          
 The mass matrix on the $(\overline{\psi}_R,\overline{\psi}_L,                  
 \overline{\chi}_R,\overline{\chi}_L) \otimes                                   
 (\psi_L,\psi_R,\chi_L,\chi_R)$ basis is                                        
\be \label{eq01}                                                                
M = \left( \begin{array}{cccc}                                                  
\mu_\psi  &  0  &  \mu_R  &  0       \\                                         
0  &  \mu_\psi  &  0  &  \mu_L       \\                                         
\mu_L  &  0  &  \mu_\chi  &  0       \\                                         
0  &  \mu_R  &  0  &  \mu_\chi                                                  
\end{array} \right) \ .                                                         
\ee                                                                             
 Here $\mu_L$, $\mu_R$ are the fermion mirror fermion mixing mass
 parameters, and the diagonal elements are produced by spontaneous              
 symmetry breaking:                                                             
\be \label{eq02}                                                                
 \mu_\psi=G_{R\psi}v_R \ , \hspace{2em} \mu_\chi=G_{R\chi}v_R \ ,               
\ee                                                                             
 with the renormalized Yukawa-couplings $G_{R\psi}$, $G_{R\chi}$ and
 the vacuum expectation value of the Higgs scalar field $v_R$.
                                                                                
 For $\mu_R \ne \mu_L$ the mass matrix $M$ in (\ref{eq01})  is not              
 symmetric, hence one has to diagonalize $M^T M$ by                             
 $O^T_{(LR)} M^T M O_{(LR)}$, and                                               
 $M M^T$ by $O^T_{(RL)} M M^T O_{(RL)}$, where                                  
\be \label{eq03}
O_{(LR)} =
           \left( \begin{array}{cccc}                                           
\cos\alpha_L  &  0  &  \sin\alpha_L  &  0   \\                                  
0  &  \cos\alpha_R  &  0  &  \sin\alpha_R   \\                                  
-\sin\alpha_L  &  0  &  \cos\alpha_L  &  0  \\                                  
0  &  -\sin\alpha_R  &  0  &  \cos\alpha_R                                      
\end{array} \right) \  ,                                                        
\ee                                                                             
 and $O_{(RL)}$ is obtained by exchanging the indices
 $R \leftrightarrow L$.
 The rotation angles of the left-handed, respectively, right-handed             
 components satisfy                                                             
\be \label{eq04}                                                                
\tan(2\alpha_L) = \frac{2(\mu_\chi \mu_L + \mu_\psi \mu_R)}                     
{\mu_\chi^2 + \mu_R^2 - \mu_\psi^2 - \mu_L^2} \ ,
\hspace{3em}
\tan(2\alpha_R) = \frac{2(\mu_\chi \mu_R + \mu_\psi \mu_L)}                     
{\mu_\chi^2 + \mu_L^2 - \mu_\psi^2 - \mu_R^2} \ ,                               
\ee                                                                             
 and the two mass-squared eigenvalues are given by                   
$$
\mu_{1,2}^2 = \half \left\{                                                     
\mu_\chi^2 + \mu_\psi^2 + \mu_L^2 + \mu_R^2
\right.
$$
\be \label{eq05}                                                                
\left.
\mp \left[ (\mu_\chi^2 - \mu_\psi^2)^2 + (\mu_L^2 - \mu_R^2)^2
+ 2(\mu_\chi^2 + \mu_\psi^2) (\mu_L^2 + \mu_R^2)                                
+ 8\mu_\chi \mu_\psi \mu_L \mu_R \right]^\half \right\} \ .
\ee                                                                             
 The mass matrix itself is diagonalized by                                      
\be \label{eq06}                                                                
O^T_{(RL)} M O_{(LR)} = O^T_{(LR)} M^T O_{(RL)}
= \left( \begin{array}{cccc}                                                    
\mu_{1}  &  0  &  0  &  0   \\                                                  
0  &  \mu_{1}  &  0  &  0   \\                                                  
0  &  0  &  \mu_{2}  &  0   \\                                                  
0  &  0  &  0  &  \mu_{2}                                                       
\end{array} \right) \ .                                                         
\ee                                                                             
 This shows that for $\mu_\psi,\mu_L,\mu_R \ll \mu_\chi$ there                  
 is a light state with mass $\mu_{1}={\cal O}(\mu_\psi,\mu_L,\mu_R)$
 and a heavy state with mass $\mu_{2}={\cal O}(\mu_\chi)$.
 In general, both the light and heavy states are mixtures of                    
 the original fermion and mirror fermion.                                       
 According to (\ref{eq04}), for $\mu_L \ne \mu_R$ the                           
 fermion-mirror-fermion mixing angle in the left-handed sector is               
 different from the one in the right-handed sector.

 In case of three mirror pairs of fermion families the diagonalization
 of the mass matrix is in principle similar but, of course, more                
 complicated.
 The strongest constraints on mixing angles between ordinary                    
 fermions and mirror fermions arise from the                                    
 conservation of $e$-, $\mu$- and $\tau$- lepton numbers and from               
 the absence of flavour changing neutral currents.
 These constraints can be avoided in a {\em ``monogamous mixing''}
 scheme, where the structure of the mass matrix is such that there is a
 one-to-one correspondence between fermions and mirror fermions.
 This happens if the family structure of the mass matrix for mirror
 fermions is closely related to the one for ordinary fermions.
                                                                                
 Let us denote doublet indices by $A=1,2$, colour indices by                    
 $c=1,2,3$ in such a way that the leptons belong to the fourth value of
 colour $c=4$, and family indices by $K=1,2,3$.                                 
 In general, the entries of the mass matrix for three mirror pairs
 of fermion families are diagonal in isospin and colour, hence they             
 have the form                                                                  
$$                                                                              
\mu_{(\psi,\chi);A_2c_2K_2,A_1c_1K_1} = \delta_{A_2A_1}                         
\delta_{c_2c_1} \mu^{(A_1c_1)}_{(\psi,\chi);K_2K_1}  \ ,                        
$$                                                                              
\be \label{eq07}                                                                
\mu_{L;A_2c_2K_2,A_1c_1K_1} = \delta_{A_2A_1}                                   
\delta_{c_2c_1} \mu^{(c_1)}_{L;K_2K_1}  \ ,
\hspace{3em}
\mu_{R;A_2c_2K_2,A_1c_1K_1} = \delta_{A_2A_1}                                   
\delta_{c_2c_1} \mu^{(A_1c_1)}_{R;K_2K_1}  \ .                                  
\ee                                                                             
 The diagonalization of the mass matrix can be achieved for given               
 indices $A$ and $c$ by two $6 \otimes 6$ unitary matrices $F_L^{(Ac)}$         
 and $F_R^{(Ac)}$ acting, respectively, on the L-handed and R-handed            
 subspaces:                                                                     
\be \label{eq08}                                                                
F_L^{(Ac)\dagger}(M^\dagger M)_L F_L^{(Ac)} \ ,
\hspace{3em}
F_R^{(Ac)\dagger}(M^\dagger M)_R F_R^{(Ac)} \ .                                 
\ee                                                                             

 The main assumption of the ``monogamous'' mixing scheme is that                
 in the family space $\mu_\psi,\mu_\chi,\mu_L,\mu_R$ are hermitian              
 and simultaneously diagonalizable, that is                                     
\be \label{eq09}                                                                
F_L^{(Ac)} = F_R^{(Ac)} =  \left(                                               
\begin{array}{cc}                                                               
F^{(Ac)}  &  0  \\  0  &  F^{(Ac)}                                              
\end{array} \right)  \ ,                                                        
\ee                                                                             
 where the block matrix is in $(\psi,\chi)$-space.                              
 The Cabibbo-Kobayashi-Maskawa matrix of quarks is given by
\be \label{eq10}                                                                
C^{(c)} \equiv F^{(2c)\dagger} F^{(1c)} \ ,
\ee                                                                             
 independently for $c=1,2,3$.                                                   
 The corresponding matrix with $c=4$ and $A=1 \leftrightarrow 2$                
 describes the mixing of neutrinos, if the Dirac-mass of the neutrinos          
 is non-zero.   
                                                                                
 A simple example for the ``monogamous'' mixing is the following:               
$$                                                                              
\mu^{(Ac)}_{\chi;K_2K_1} = \lambda^{(Ac)}_\chi                                  
\mu^{(Ac)}_{\psi;K_2K_1} + \delta_{K_2K_1}\Delta^{(Ac)} \ ,                     
$$                                                                              
\be \label{eq11}                                                                
\mu^{(c)}_{L;K_2K_1} = \delta_{K_2K_1}\delta^{(c)}_L \ ,
\hspace{3em}
\mu^{(Ac)}_{R;K_2K_1} = \lambda^{(Ac)}_R                                        
\mu^{(Ac)}_{\psi;K_2K_1} + \delta_{K_2K_1}\delta^{(Ac)}_R \ ,                   
\ee                                                                             
 where, as the notation shows, $\lambda_\chi^{(Ac)},\; \Delta^{(Ac)},\;
 \delta_L^{(c)},\; \lambda_R^{(Ac)},\; \delta_R^{(Ac)}$ do not depend
 on the family index.

 The full diagonalization of the mass matrix on the                             
 $(\psi_L,\psi_R,\chi_L,\chi_R)$ basis of all three family pairs is             
 achieved by the $96 \otimes 96$ matrix                                         
\be \label{eq12}                                                                
{\cal O}^{(LR)}_{A^\prime c^\prime K^\prime,AcK}                                
= \delta_{A^\prime A}\delta_{c^\prime c}                                        
F^{(Ac)}_{K^\prime K}
\cdot \left(                                                                    
\begin{array}{cc}                                                               
 \cos\alpha^{(AcK)}_L  &  0  \\                                                 
 0  &  \cos\alpha^{(AcK)}_R  \\                                                 
-\sin\alpha^{(AcK)}_L  &  0  \\                                                 
 0  & -\sin\alpha^{(AcK)}_R                                                     
\end{array}
\begin{array}{cc}                                                               
 \sin\alpha^{(AcK)}_L  &  0  \\                                                 
 0  &  \sin\alpha^{(AcK)}_R  \\                                                 
 \cos\alpha^{(AcK)}_L  &  0  \\                                                 
 0  &  \cos\alpha^{(AcK)}_R                                                     
\end{array}  \right) \ .                                                        
\ee                                                                             
 ${\cal O}^{(RL)}$ is obtained from ${\cal O}^{(LR)}$ by                  
 $\alpha_L \leftrightarrow \alpha_R$.                                           
                                                                                
 In case of $\mu_R=\mu_L$ the left-handed and right-handed mixing
 angles are the same:                                       
\be \label{eq13}                                                                
\alpha^{(AcK)} \equiv \alpha_L^{(AcK)} =                                        
\alpha_R^{(AcK)} \ .                                                            
\ee                                                                             
 In ref.~\cite{FAMILIES} only this special case was considered.
 The importance of the left-right-asymmetric mixing was pointed out in
 ref.~\cite{CSIFOD}, where the constraints arising from the measured            
 values of anomalous magnetic moments were investigated.
 It turned out that strong constraints arise for the product
 $\alpha^{(l)}_L \alpha^{(l)}_R$ ($l=e,\mu$), but the individual
 values $\alpha^{(l)}_L$ and $\alpha^{(l)}_R$ are much less
 restricted.
 In case of the L-R asymmetric mixing these constraints can be
 satisfied, if either the left- or right-handed mixing exactly           
 vanishes (or is very small): $\alpha^{(l)}_L \simeq 0$ or
 $\alpha^{(l)}_R \simeq 0$.

 Another important observation in ref.~\cite{CSIFOD} is the relation
\be \label{eq14}                                                                
\mu^{(c)}_L = -\sin\alpha^{(AcK)}_R \cos\alpha^{(AcK)}_L \mu^{(AcK)}_1
+\sin\alpha^{(AcK)}_L \cos\alpha^{(AcK)}_R \mu^{(AcK)}_2 \ ,
\ee                                                                             
 which implies for small mixing angles and large mass differences
 between the two mixed fermion states
\be \label{eq15}                                                                
\mu^{(c)}_L \simeq \alpha^{(AcK)}_L \mu^{(AcK)}_2 
\hspace{3em} (A=1,2;\; K=1,2,3)  \ .
\ee                                                                             
 Hence the left-handed mixing angles are inversely proportional to
 the heavy fermion masses.
 In addition, in case of the simple mass matrix pattern in (\ref{eq11}),
 the left-hand side is independent of the isospin and family index.
 No such constraints exist for the right handed mixing angles, therefore
 the choice
\be \label{eq16}                                                                
\mu^{(c)}_L,\; \alpha^{(AcK)}_L \simeq 0 \ ,
\hspace{3em} |\alpha^{(AcK)}_L| \ll |\alpha^{(AcK)}_R|
\ee                                                                             
 leaves more freedom for large mixing than the other possibility,
 namely, $\mu^{(Ac)}_R,\; \alpha^{(AcK)}_R \simeq 0$.

\section{Explanation of $R_b$ and $R_c$}\label{sec3}
 In order to discuss the couplings of the electroweak vector bosons
 ($A,W,Z$), let us consider the form of the electroweak currents.
 The electroweak interaction can be written in general as
\be \label{eq17}                                                                
eJ_Q(x)_\mu A(x)_\mu 
+ g\left[ J^-_L(x)_\mu W^+(x)_\mu +  J^+_L(x)_\mu W^-(x)_\mu \right]
+ \frac{e}{\sin\theta_W \cos\theta_W}J_Z(x)_\mu Z(x)_\mu  \ .
\ee                                                                             
 Here $\theta_W$ is the Weinberg-angle 
 ($\sin\theta_W \equiv e/g,\; \sin^2\theta_W \simeq 0.225$).

 Let us denote the fermion fields corresponding to the mass eigenstates
 by $\xi^{(AcK)}(x)$ for the light states and  $\eta^{(AcK)}(x)$ for the
 heavy states, respectively.
 These are linear combinations of the fermion ($\psi^{(AcK)}(x)$) and
 mirror fermion ($\chi^{(AcK)}(x)$) fields, as discussed in the previous
 section.
 The electromagnetic current $J_Q(x)_\mu$ in (\ref{eq17}) has the
 same form in terms of $\xi$ and $\eta$ as in terms of $\psi$ and
 $\chi$:
\be \label{eq18}                                                                
J_Q(x)_\mu = \sum_{A,c,K} Q^{(Ac)} \left\{
\overline{\xi}^{(AcK)}(x) \gamma_\mu \xi^{(AcK)}(x) +
\overline{\eta}^{(AcK)}(x) \gamma_\mu \eta^{(AcK)}(x) \right\} \ ,
\ee                                                                             
 where $Q^{(Ac)}$ denotes the electromagnetic charge.
 The neutral current $J_Z(x)_\mu$ can be expressed as
\be \label{eq19}                                                                
J_Z(x)_\mu = \sin^2\theta_W J_Q(x)_\mu - J_{L3}(x)_\mu \ ,
\ee                                                                             
 with the third isospin component of the left-handed current
 $J_{L3}(x)_\mu$.
 This latter depends on the mixing angles but is still diagonal in
 isospin:
$$
J_{L3}(x)_\mu = \half\sum_{A,c,K} (-1)^{1+A} \left\{
\cos^2\alpha^{(AcK)}_L
\overline{\xi}^{(AcK)}_L(x) \gamma_\mu \xi^{(AcK)}_L(x) 
+ \sin^2\alpha^{(AcK)}_R
\overline{\xi}^{(AcK)}_R(x) \gamma_\mu \xi^{(AcK)}_R(x)
\right.
$$
$$
+ \sin^2\alpha^{(AcK)}_L
\overline{\eta}^{(AcK)}_L(x) \gamma_\mu \eta^{(AcK)}_L(x)
+ \cos^2\alpha^{(AcK)}_R
\overline{\eta}^{(AcK)}_R(x) \gamma_\mu \eta^{(AcK)}_R(x)
$$
$$
+ \cos\alpha^{(AcK)}_L \sin\alpha^{(AcK)}_L \left[
\overline{\xi}^{(AcK)}_L(x) \gamma_\mu \eta^{(AcK)}_L(x) +
\overline{\eta}^{(AcK)}_L(x) \gamma_\mu \xi^{(AcK)}_L(x) \right]
$$
\be \label{eq20} 
\left.                                                               
- \cos\alpha^{(AcK)}_R \sin\alpha^{(AcK)}_R \left[
\overline{\xi}^{(AcK)}_R(x) \gamma_\mu \eta^{(AcK)}_R(x) +
\overline{\eta}^{(AcK)}_R(x) \gamma_\mu \xi^{(AcK)}_R(x) \right]
\right\} \ .
\ee                                                                             
 Finally, the charged current contains also the CKM-matrices
 defined in (\ref{eq10}):
$$
J_L^+(x)_\mu = \frac{1}{\sqrt{2}} \sum_{c,K_1,K_2} 
C^{(c)}_{K_1K_2} \cdot
$$
$$
\cdot \left\{
\cos\alpha^{(1cK_1)}_L \cos\alpha^{(2cK_2)}_L
\overline{\xi}^{(2cK_2)}_L(x) \gamma_\mu \xi^{(1cK_1)}_L(x) 
+ \sin\alpha^{(1cK_1)}_R \sin\alpha^{(2cK_2)}_R
\overline{\xi}^{(2cK_2)}_R(x) \gamma_\mu \xi^{(1cK_1)}_R(x)
\right.
$$
$$
+ \sin\alpha^{(1cK_1)}_L \sin\alpha^{(2cK_2)}_L
\overline{\eta}^{(2cK_2)}_L(x) \gamma_\mu \eta^{(1cK_1)}_L(x)
+ \cos\alpha^{(1cK_1)}_R \cos\alpha^{(2cK_2)}_R
\overline{\eta}^{(2cK_2)}_R(x) \gamma_\mu \eta^{(1cK_1)}_R(x)
$$
$$
+ \cos\alpha^{(1cK_1)}_L \sin\alpha^{(2cK_2)}_L 
\overline{\eta}^{(2cK_2)}_L(x) \gamma_\mu \xi^{(1cK_1)}_L(x)
+ \sin\alpha^{(1cK_1)}_L \cos\alpha^{(2cK_2)}_L 
\overline{\xi}^{(2cK_2)}_L(x) \gamma_\mu \eta^{(1cK_1)}_L(x) 
$$
\be \label{eq21} 
\left.                                                               
- \sin\alpha^{(1cK_1)}_R \cos\alpha^{(2cK_2)}_R 
\overline{\eta}^{(2cK_2)}_R(x) \gamma_\mu \xi^{(1cK_1)}_R(x)
- \cos\alpha^{(1cK_1)}_R \sin\alpha^{(2cK_2)}_R 
\overline{\xi}^{(2cK_2)}_R(x) \gamma_\mu \eta^{(1cK_1)}_R(x) 
\right\} \ .
\ee                                                                             
 Note that the CKM-structure in the charged currents is exactly
 conserved only in the case of
\be \label{eq22} 
\alpha^{(AcK_1)}_L = \alpha^{(AcK_2)}_L  \ , 
\hspace{3em}
\alpha^{(AcK_1)}_R = \alpha^{(AcK_2)}_R  \ ,
\ee                                                                             
 that is for universal mixing angles independent of the family index.
 In this interesting special case both charged and neutral currents
 satisfy exact universality with respect to the family index, if quarks
 and leptons are considered separately.

 There is, however, still a possibility to break universality of the
 couplings between quarks and leptons.
 Such universality relations are not as strong as for leptons and
 quarks separately, because there the strong coupling $\alpha_s$
 appears due to QCD radiative corrections.
 For instance, in the Z-boson widths the QCD correction factors for
 hadronic final states originating from light ($m_q \simeq 0$) quarks
 have the form $1+K_{QCD}$ with \cite{QCD}
\be \label{eq23} 
K_{QCD} = \frac{\alpha_s}{\pi} 
+ 1.409 \left( \frac{\alpha_s}{\pi} \right)^2
- 12.77 \left( \frac{\alpha_s}{\pi} \right)^3 + \ldots \ .
\ee                                                                             
 Therefore the existing few percent uncertainty in $\alpha_s(M_Z)$
 results in an overall uncertainty of the order of percents in the
 universality relation of the hadronic versus leptonic $Z$ decay
 widths.

 Another observation about the currents in eqs.~(\ref{eq20}-\ref{eq21})
 is that on the light states, for small mixing angles, the admixtures
 of $(V+A)$-type are proportional to $\alpha_R^2$ (in amplitude).
 This means that both neutral and charged currents are strongly
 dominated by the $(V-A)$ couplings, in accordance with observations.
 Moreover, choosing the ordering in (\ref{eq16}), and in addition
 requiring no (or very small) mixings for one of the members of
 the isospin doublets, for instance,
\be \label{eq24} 
\alpha^{(\nu)}_R, \alpha^{(d)}_R \simeq 0 \ ,
\ee                                                                             
 the charged currents on the light states become perfectly $(V-A)$.

 After these preparations let us now turn in more detail to the
 question of violations of universality in Z-boson decay. 
 According to eqs.~(\ref{eq19}-\ref{eq20}) the usual coupling
 parameters $g_V^f$ and $g_A^f$ for the vector and axialvector
 couplings of the Z-boson become
$$
g_V^f = T_{3L}\left[ \cos^2\alpha_L^{(f)}+\sin^2\alpha_R^{(f)} \right]
- 2Q_f\sin^2\theta_W \ , 
$$
\be \label{eq25} 
g_A^f = T_{3L}\left[ \cos^2\alpha_L^{(f)}-\sin^2\alpha_R^{(f)} \right]
\ ,
\ee                                                                             
 where $T_{3L}=\pm \half$ is the third component of the left-handed
 isospin, $\alpha_{L,R}^{(f)} \equiv \alpha_{L,R}^{(AcK)}$ denotes
 the mixing angles and $Q_f \equiv Q^{(Ac)}$ is the electric charge in
 units of the positron charge.
 Up to order ${\cal O}(\alpha^2_{L,R})$, we obtain the following
 tree-level corrections for the $Z \to ff$ widths:
$$
\frac{\Gamma_{Z \to ff}}{\Gamma^{(sm)}_{Z \to ff}} =
\frac{\cos^4\alpha_L^{(f)} + \sin^4\alpha_R^{(f)}
- 4s_W^2|Q_f| (\cos^2\alpha_L^{(f)} + \sin^2\alpha_R^{(f)}) 
+ 8s_W^4 Q_f^2}{1 - 4s_W^2|Q_f| + 8s_W^4 Q_f^2}
$$
\be \label{eq26} 
= \frac{1 - 4s_W^2|Q_f| + 8s_W^4 Q_f^2
- \alpha_L^{(f)2}(2-4s_W^2|Q_f|) - \alpha_R^{(f)2}4s_W^2|Q_f|}
{1 - 4s_W^2|Q_f| + 8s_W^4 Q_f^2} + {\cal O}(\alpha^4_{L,R}) \ ,
\ee                                                                             
 where the index $(sm)$ denotes the standard model expressions and
 $s_W^2 \equiv \sin^2\theta_W$.
 Similarly, the corrections for the usual parameters in the
 forward-backward asymmetries 
\be \label{eq27} 
A_f \equiv \frac{2g_V^f g_A^f}{(g_V^f)^2+(g_A^f)^2}
\ee                                                                             
 have the form
$$
\frac{A_f}{A_f^{(sm)}} =
\frac{\cos^4\alpha_L^{(f)} - \sin^4\alpha_R^{(f)}
- 4s_W^2|Q_f| (\cos^2\alpha_L^{(f)} - \sin^2\alpha_R^{(f)})}
{\cos^4\alpha_L^{(f)} + \sin^4\alpha_R^{(f)}
- 4s_W^2|Q_f| (\cos^2\alpha_L^{(f)} + \sin^2\alpha_R^{(f)}) 
+ 8s_W^4 Q_f^2}
$$

$$
= 1 + \alpha_L^{(f)2} (2-4s_W^2|Q_f|) 
\left[ (1 - 4s_W^2|Q_f| + 8s_W^4 Q_f^2)^{-1} 
- (1 - 4s_W^2|Q_f|)^{-1} \right]
$$

\be \label{eq28} 
+ \alpha_R^{(f)2} 4s_W^2|Q_f|
\left[ (1 - 4s_W^2|Q_f| + 8s_W^4 Q_f^2)^{-1} 
+ (1 - 4s_W^2|Q_f|)^{-1} \right] + {\cal O}(\alpha^4_{L,R}) \ .
\ee                                                                             

 The comparison of these expressions to the high precision leptonic
 data \cite{LEP} shows that the leptonic mixing angles are small: 
 $\alpha_{L,R}^{(e,\mu,\tau,\nu)} 
 = {\cal O}(10^{-3})-{\cal O}(10^{-4})$.
 This is in agreement with earlier observations, for instance, in
 ref.~\cite{CSIFOD}.
 In the hadronic sector let us concentrate on the experimental data 
 \cite{LEP} $\Gamma_{had}=1744.8(3.0)$ MeV, $R_b=0.2219(17)$,
 $R_c=0.1543(74)$ and $A_{FB}^{(0,b)}/A_{FB}^{(0,c)}=A_b/A_c=1.38(15)$.
 In the absence of a full one-loop calculation, we have to use the
 tree-level expressions.
 For the Weinberg-angle we consider the values $s_W^2=0.230$ and
 $s_W^2=0.225$.
 The former is a typical value obtained by the one-loop fits within
 the minimal standard model \cite{LEP}, whereas the latter is obtained
 from the tree-level relation $s_W^2=1-M_W^2/M_Z^2$. 
 For the strong coupling at $Q^2=M_Z^2$ we take the three
 representative values $\alpha_s=0.123,\; 0.130,\; 0.116$.
 Assuming that the observed deviations from universality in $R_b$ and
 $R_c$ are due to a single dominant mixing angle parameter, I tried
 fits of the above data in several different settings.
 Representative examples are:
\be \label{eq29} 
A:\; \alpha_R^{(c)2} \ne 0 \ ,  \hspace{2em}
B:\; \alpha_L^{(u)2}=\alpha_L^{(d)2}
=\alpha_L^{(c)2}=\alpha_L^{(s)2} \ne 0 \ ,  \hspace{2em}
C:\; \alpha_L^{(c)2}=\alpha_L^{(s)2} \ne 0 \ .
\ee                                                                             
 In case of $s_W^2=0.230$, $\alpha_s=0.123 \pm 0.007$ the obtained
 $\chi^2$ values were in the range $\chi^2=6-7$ and the best fits for
 the mixing angles were compatible with zero.
 (Possible correlations in the data are neglected here.)
 This reflects the well known fact that the chosen set of data cannot
 be well described by the minimal standard model.
 (In fact, taking into account the one-loop electroweak corrections
 makes the fit even worse, see ref.~\cite{LEP}.)
 In cases $B$ and $C$ better fits are possible with the tree-level
 value $s_W^2=0.225$ and $\alpha_s=0.130$: for case $B$ 
 $\alpha_L^{(u)2}=\alpha_L^{(d)2}=\alpha_L^{(c)2}=\alpha_L^{(s)2}=
 0.0054(10),\; \chi^2=4.7$, and for case $C$
 $\alpha_L^{(c)2}=\alpha_L^{(s)2}=0.011(2),\; \chi^2=3.5$.
 In these cases the other values of $\alpha_s$ give slightly worse
 $\chi^2$.
 In case $A$ $s_W^2=0.225$ leads to substantially worse fits than
 $s_W^2=0.230$, namely $\chi^2=9-12$.
 This shows that the best case is:
\be \label{eq30} 
C:\; \alpha_L^{(c)2}=\alpha_L^{(s)2} =0.011(2) \ ,
\hspace{3em} (\chi^2=3.5)  \ .
\ee                                                                             
 The best fit for case $B$ is somewhat worse than for $C$, but still
 better than the standard model fit.

\section{Discussion}\label{sec4}
 As it has been shown in the previous section, the violations of
 universality in $R_b$ and $R_c$ at LEP can be explained in a
 vectorlike extension of the standard model by the mixing of the
 low-lying fermion states with the heavy mirror fermion partners.
 A simple choice of the dominant mixing angles for this is given in
 eq.~(\ref{eq30}), where the universality violations are mainly due
 to the equal left-handed mixings of the second family quarks. 
 By this choice the violations of universality at tree level are
 minimized but, of course, other mixing schemes with several
 non-negligible mixing angles cannot be excluded at present.
 (For some other ways to explain $R_b$ and $R_c$ by mixing to
 heavy states see ref.~\cite{MIXING}.)

 The analysis in the present paper relies on the tree approximation.
 For a more accurate treatment a full calculation of one-loop
 radiative corrections is necessary, including also vertex- and
 box-corrections.
 First steps in this direction were done in ref.~\cite{CSCSI}, where
 the one-loop effects in vector boson propagators were considered and
 an overall fit of the experimental data was performed, taking into
 account radiative corrections.

 The better description of the hadronic branching ratios of the
 Z-boson is a hint for the vectorlike extension of the standard model
 considered here.
 The final check is, of course, to find the heavy mirror fermions.


\end{document}